# A Generic and Automated Methodology to Simulate Melting Point

Fu-Zhi Dai [a, b], Si-Hao Yuan [a], Yan-Bo Hao [a], Xin-Fu Gu [a] *, Shipeng Zhu [c] **,

Jidong Hu [c], Yifen Xu [c]

[a] School of Materials Science and Engineering, University of Science and Technology Beijing, Beijing, 100083, P. R. China

[b] AI for Science Institute, Beijing, 100084, P. R. China

[c] Science and Technology on Advanced Functional Composite Laboratory, Aerospace Research Institute of Materials & Processing Technology, Beijing 100076, P. R. China

Corresponding author.

* X.-F. Gu: E-mail: (xinfugu@ustb.edu.cn)

** S. Zhu: E-mail: (carbonfiber703@163.com)

**Abstract**

The melting point of a material constitutes a pivotal property with profound implications across various disciplines of science, engineering, and technology. Recent advancements in machine learning potentials have revolutionized the field, enabling ab initio predictions of materials' melting points through atomic-scale simulations. However, a universal simulation methodology that can be universally applied to any material remains elusive. In this paper, we present a generic, fully automated workflow designed to predict the melting points of materials utilizing molecular dynamics simulations. This workflow incorporates two tailored simulation modalities, each addressing scenarios with and without elemental partitioning between solid and liquid phases. When the compositions of both phases remain unchanged upon melting or solidification, signifying the absence of partitioning, the melting point is identified as the temperature at which these phases coexist in equilibrium. Conversely, in cases where elemental partitioning occurs, our workflow estimates both the nominal melting point, marking the initial transition from solid to liquid, and the nominal solidification point, indicating the reverse process. To ensure precision in determining these critical temperatures, we employ an innovative temperature-volume data fitting technique, suitable for a diverse range of materials exhibiting notable volume disparities between their solid and liquid states. This comprehensive approach offers a robust and versatile solution for predicting melting points, fostering advancements in materials science and technology.

# 1. Introduction

For years, the aspiration to digitally design materials has fueled countless innovations, and this dream is now closer to reality than ever before. With the relentless advancements in computational capabilities and tools, coupled with the soaring progress in artificial intelligence (AI), we are entering an era where predicting material properties and generating novel materials is not only feasible but increasingly precise and efficient. Specifically, the evolution of machine learning potentials has resolved the long-standing conflict between efficiency and accuracy in atomic-scale simulations, paving the way for remarkable enhancements in both materials property prediction and the generation of new materials [1–7]. To fully harness the capacity of machine learning potentials in predicting new materials, we necessitate generic and automated workflows to simulate materials properties.

The melting point carries profound implications across diverse domains of science, engineering, and technology, *e.g.* searching for thermal protection materials in aerospace [7,8]. The melting point, unlike material quantities like elastic or lattice constants, is computationally challenging. In atomic-scale simulations, it's determined by finding the temperature where the Gibbs free energies of the solid and liquid phases equalize at constant pressure. There are two prevalent methods exist for calculating this:

(1) The Free Energy Approach [9]: This involves calculating the Gibbs free energies of both phases explicitly, often through thermodynamic integration. The melting point is identified where these energies intersect.

(2) The Interface Method (or Coexistence Approach) [10]: It simulates the coexistence of solid and liquid phases to estimate the melting point.

Several computational tools have been devised, grounded in these principles, to facilitate the computation of melting points [11–13]. Nevertheless, these tools still face numerous constraints in accurately addressing diverse materials:

(1) The tool created by Zhu et al. [11], which employs molecular dynamics (MD) simulations and a coexistence approach to predict melting points, is confined to simple elemental crystals.

(2) Both the methods developed by Hong et al. [12] and Klimanova et al. [13] share a similar underlying philosophy and, in theory, can be applied to any material. However, they are optimized for smaller simulation supercells, such as those used in first-principles simulations, and do not fully leverage the strengths of MD simulations with larger supercells.

(3) The most critical limitation of prior methods lies in their neglect of elemental partitioning. When the solid and liquid phases in equilibrium share the same composition, these methods yield accurate predictions. However, when there is a difference in composition between the two phases, elemental partitioning occurs, resulting in a melting range rather than a single melting point. The lower limit corresponds to the temperature at which the solid phase begins to melt, while the upper limit marks the temperature at which the liquid phase starts to solidify. Consequently, the melting points predicted by these methods have limited practical applicability, since most real materials exhibit elemental partitioning during melting, *e.g.* solid solutions or even high entropy materials.

To overcome these limitations and fully harness the capability of machine learning potentials in predicting material properties, we have devised a generic and automated approach for calculating material melting points. This methodology is suitable for materials that exhibit or do not exhibit elemental partitioning during melting. For materials without partitioning, our method predicts a single melting point. Conversely, for materials that undergo partitioning, we can compute both the initial melting point of the solid phase and the initial solidification point of the liquid phase. Furthermore, our approach is applicable to any material where there is a significant difference in the volumes of the solid and liquid phases.

**2. Methodology**

The overall methodology we employ encompasses two pivotal stages, as illustrated in Fig. 1a: the fast-scanning stage and the iterative refinement stage.

(1) In the fast-scanning stage, a fully solid-state material (as depicted in Fig. 1b) undergoes a gradual temperature increase, starting from a low value and reaching a peak temperature high enough to ensure complete liquefaction. With this melting process, we can obtain a rough estimation of the melting temperature $T_m^0$. It is used as the upper boundary for subsequent simulations, accounting for the overheating phenomenon inherent to the scanning process. Moreover, it can also be adopted as a reference temperature to generate the solid-liquid two-phase structure.

(2) Subsequently, the iterative refinement stage commences, where the material is equilibrated across a range of temperatures, initiating from a solid-liquid two-phase structure (a mixed state, Fig. 1c and Fig. 1d). For the initial iteration, the temperature bounds are set at $T_{min} = T_m^0/2$ and $T_{max} = T_m^0$, with a temperature step of $\Delta T = T_m^0/20$. In subsequent iterations, the temperature range adjusts dynamically based on the estimated melting temperature from the previous iteration, $T_m^{i-1}$, such that $T_{min} = T_m^{i-1} - \Delta T^{i-1}$ and $T_{max} = T_m^{i-1} + \Delta T^{i-1}$. Here, $T_m^{i-1}$ represents the refined melting temperature estimation after finishing the iteration step of *i*-1. Notably, the temperature step size $\Delta T$ decreases with each iteration, following the pattern $\Delta T^i = \Delta T^{i-1}/5$, allowing for increasingly precise refinement of the melting temperature. When $\Delta T^i$ is lower than a threshold, the iteration stage stops.

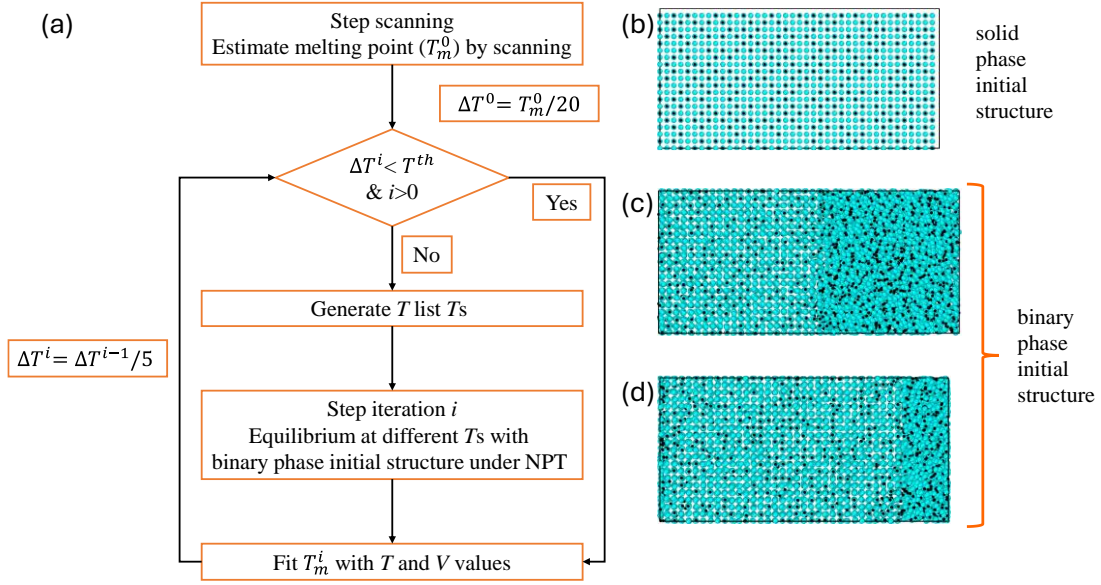

Figure 1. Illustration of the automate melting point simulation process, along with various states of a HfC structure. (a) Flowchart outlining the steps involved in the simulation process. (b) A fully solid state of the HfC structure. (c) A mixed state with half of the HfC structure in liquid phase. (d) A mixed state with a small amount of liquid present.

The accurate estimation of $T_m^i$ is central to ensuring the seamless operation of our methodology. Contrary to prior research, we have devised a novel approach to estimating $T_m^i$ by leveraging readily accessible $T\sim V$ (temperature versus volume) data. This data was compiled from extensive simulations and subsequently fitted to a sigmoid-like function, facilitating a precise representation of the temperature-induced volume changes. The fitted function, given by:

$$V = \frac{b}{1+\exp(-a(T-T_0))} + cT + V_0 \qquad (1)$$

elegantly captures the nuances of the material's response to temperature changes. Here, $a$, $b$, $c$, $T_0$, and $V_0$ are fitting parameters. Specifically, the term $cT + V_0$ accounts for the linear expansion of the material with increasing temperature, while the sigmoid function $\frac{b}{1+\exp(-a(T-T_0))}$ precisely depicts the abrupt volume jump during the melting process. Notably, $T_0$ in this equation represents the melting point of the material.

In our methodology, we conduct all molecular dynamics (MD) simulations within the isothermal-isobaric ensemble (NPT). All the simulations are carried out by using the LAMMPS code [14]. For each simulation, we adopt a timestep of 1 femtosecond (fs) to ensure accurate time evolution of the system. The damping parameters are specifically set to 500 fs for pressure and 100 fs for temperature, enabling effective control over these thermodynamic variables. The input structure required for simulations is a fully solid-state configuration, akin to the one depicted in Fig. 1b. In the fast-scanning stage, the entire process comprises a total of 1,000,000 simulation steps with temperature gradually increased from $T_{low}$ to $T_{high}$. Importantly, in the iteration stage, our methodology is capable of autonomously generating solid-liquid coexistent structures (mixed states, as shown in Fig. 1c and Fig. 1d). These mixed states feature a solid-liquid interface located in the *xy* plane, necessitating a dimension in the *z*-direction that is typically twice that of the *x*- and *y*-directions, while the latter two maintain similar dimensions. The simulation protocol for creating these mixed states comprises two main stages. First, a rapid heating spans 50,000 steps, wherein the system is gradually heated to a temperature *T*, which is set below the upper limit $T_m^0$ to ensure the system keeps in solid state. Subsequently, during another 50,000 steps, we selectively manipulate the system to generate the desired mixed state. This involves either freezing half of the system (leading to a structure resembling Fig. 1c) or a majority portion (resulting in a configuration akin to Fig. 1d), while the unfrozen portion is heated above $T_m^0$ to induce liquid formation. Finally, to ensure that the generated mixed state reaches thermodynamic equilibrium, we perform an equilibration stage that lasts for 500,000 steps. During this period, the system is maintained at the set temperature *T* and pressure (usually 0 Pa, although we emphasize that it is generalizable to arbitrary-pressure calculations).

### 3. Results and Discussion
### 3.1 Simple systems without elemental partitioning

The melting point of a material is clearly defined when the composition of the solid and the liquid phase remains identical, indicating an absence of elemental partitioning during the melting or solidification process. At the melting point, the Gibbs free energy of the solid and liquid phase is the same, or alternatively the probability of the existence of the solid and the liquid phase is equal, *i.e.*, $P_{\text{solid}}/P_{\text{liquid}}$ = 1. In the realm of atomic-scale simulations, the solid-liquid coexistence method harnesses this principle by seeking out an equilibrium mixed state characterized by a half-solid, half-liquid configuration or to search for the condition that the two phase have the same probability to exist [11–13].

Our methodology also embraces this concept by simulating the melting point. To exemplify this, let's consider the estimation of aluminum's melting point. Fig. 2 meticulously depicts the estimation procedure in our workflow. Fig. 2a showcases the initial temperature-volume (*T~V*) data acquired during the fast-scanning stage, where a pronounced volume jump signifies the transition from a fully solid to a fully liquid state. This *T~V* data is seamlessly fitted to Equation (1), with $T_0$ ($T_m^0$=1126 K) strategically positioned near the midpoint of the volume jump, capturing the essence of the melting event. Then, the melting point undergoes refinement in the subsequent iteration stage. In this iterative process, akin to the conventional solid-liquid coexistence simulation technique, half of the material is initially melted to establish the starting condition for each simulation task, as depicted in Fig. 1c. Fig. 2b to Fig. 2d sequentially illustrate the refinement and convergence of the predicted melting point value through subsequent iterative stages. Specifically, the melting point estimates progressively adjust to $T_m^1$=947 K, $T_m^2$=932 K, and finally settle at $T_m^3$=925 K. This sequence underscores the remarkable fitting accuracy of Equation (1) in capturing the intricate temperature-dependent volume behavior, pinpointing the critical melting point as the precise midpoint of the volume jump. Fig. 2e compiles the $T_{min}$, $T_{max}$, and intermediate melting point estimates ($T_m^i$) encountered during the simulations, plotting their evolution against the iteration count (with 0 representing the fast-scanning stage). This visualization underscores both the convergence of the iterative fitting process and the robustness of the final melting point estimate, highlighting the stability and reliability of our approach.

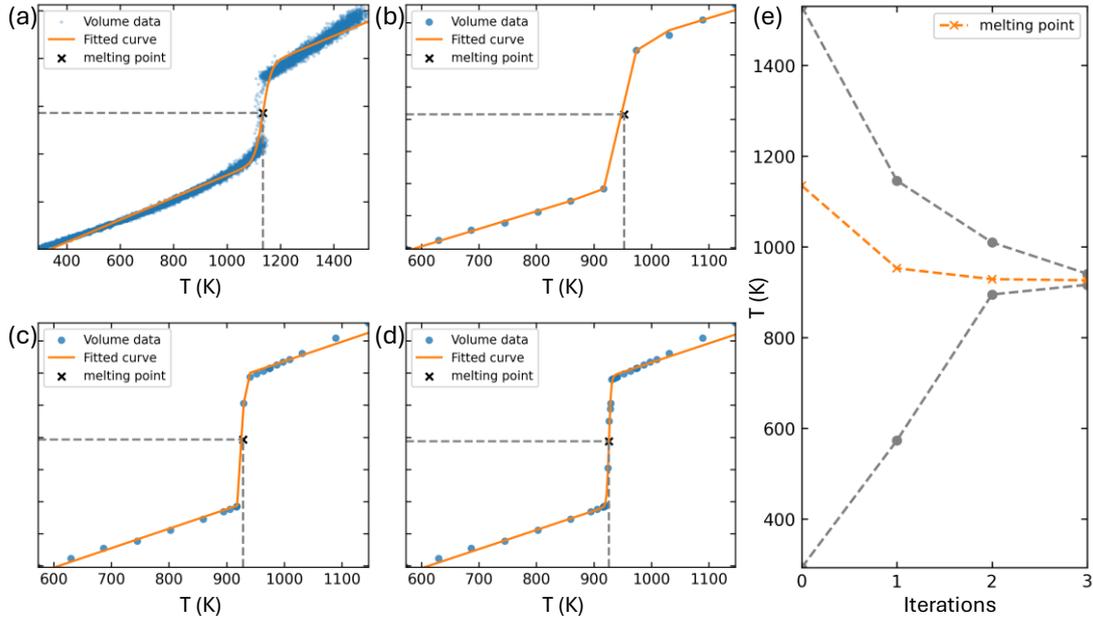

Figure 2. Fitting the *T~V* (temperature versus volume) data of pure Al to estimate the melting point through iterative processes. (a) The initial *T~V* data obtained from the fast-scanning process. The refined *T~V* data after (b) the first iteration, (c) the second iteration and (d) the third iteration. (e) Variation of the estimated melting point, the highest simulating temperature, and the lowest simulating temperature with respect to the number of iterations, illustrating the convergence of the iterative fitting process and the stability of the final melting point estimate.

To demonstrate the effectiveness of our approach, we simulated the melting points of various materials, including Aluminum (FCC), Magnesium (HCP), Iron (BCC), and Silicon (Diamond structure). The potential model adopted are the same as those provided in Zhu's software [11]. The outcomes of these simulations are presented in Table 1, where we also compare them against the predicted values from Zhu's work [11]. It reveals a good agreement between our results and Zhu's predictions, underscoring the accuracy of our methodology. This alignment not only attests to the correctness of our approach but also highlights its potential to accurately predict melting points across diverse material structures with only *T~V* data.

Table 1. Comparison of melting point predicted by our method and the values predicted by Zhu's method [11]. The interatomic potentials for Al is the Finnis–Sinclair type embedded atom method (EAM) potential developed by Mendelev et al. [15], for Mg is the EAM potential developed by [16], for Fe is the EAM potential developed by Becquart et al. [17], and for Si is the Stillinger-Weber potential developed by Stillinger and Weber [18].

|  | Al (FCC) | Mg (HCP) | Fe (BCC) | Si (Diamond) |
| --- | --- | --- | --- | --- |
| This work | 926 K | 912 K | 1795 K | 1677 K |
| Zhu's work [11]* | 926 K | 911 K | 1790 K | 1665 K |

* The data are from the github: https://github.com/pyiron/pyiron_meltingpoint

### 3.2 Generic systems with elemental partitioning

Typically, the melting point of a material is not a sharply defined value but rather a range, as depicted in Fig. 3, which illustrates a binary phase diagram featuring a solidus and a liquidus line. When the material's temperature exceeds the liquidus line, it exists in a fully liquid state. Conversely, if its temperature falls below the solidus line, it is fully solid. In these extreme conditions, the composition of both the solid and liquid phases is uniform. However, when the temperature lies between the solidus and liquidus lines, the material exists in a state of solid-liquid coexistence. In this intermediate zone, the compositions of the solid and liquid phases diverge from the average composition, indicating the occurrence of elemental partitioning between these two phases. For a material with a specific composition, as it is heated from the fully solid state towards the solidus line, melting commences. For the sake of simplicity, we refer to the intersection point between the solidus line and the composition line as the nominal melting point. Conversely, when the material is cooled from the fully liquid state towards the liquidus line, solidification begins. Similarly, the intersection of the liquidus line with the composition line is referred to as the nominal solidification point for simplicity.

When the temperature falls slightly below the solidus line, the material attains a fully solid equilibrium state, as depicted in Fig. 3c. Conversely, a slight elevation above the solidus line results in a mixed equilibrium state, characterized by a minor proportion of liquid, as illustrated in Fig. 3b. Notably, the composition of the solid phase in this state closely aligns with the average composition of the material. On the other hand, if the temperature surpasses the liquidus line by a marginal amount, the material transitions to a fully liquid equilibrium state, as shown in Fig.3d. When the temperature drops slightly below the liquidus line, a mixed state emerges, featuring a small fraction of solid, as evidenced in Fig. 3e. Importantly, the composition of the liquid phase in this mixed state approximates the overall average composition of the material.

At the nominal solidification point, precisely at the liquidus line, the equilibrium state is a delicate balance of predominantly liquid with an infinitesimal amount of solid, and the liquid phase composition coincides with the average composition. Similarly, at the nominal melting point, which aligns with the solidus line, the equilibrium state comprises a negligible amount of liquid within a primarily solid matrix, where the solid phase composition mirrors the average composition. These inherent properties will serve as the cornerstone for estimating both the melting and solidification points within our methodology.

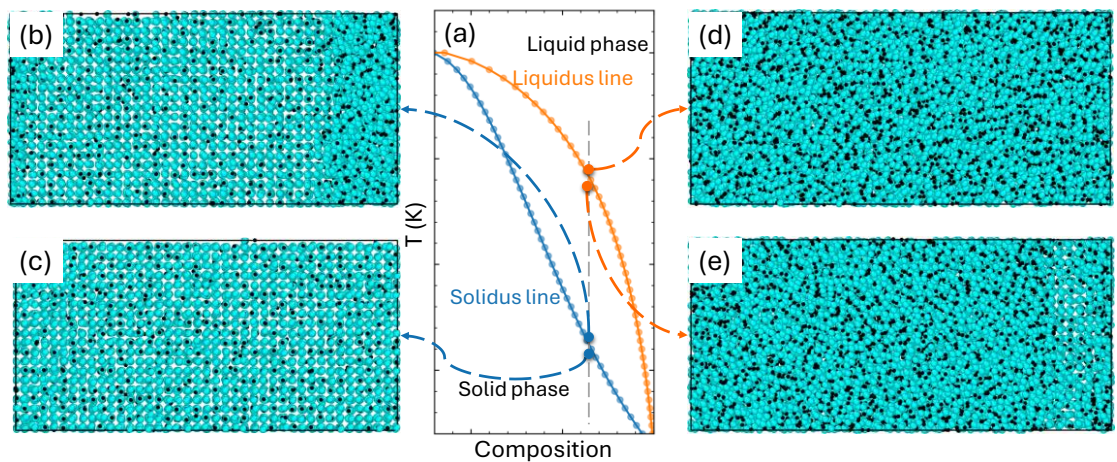

Figure 3. (a) Schematic illustration of a binary phase diagram. The diagram illustrates the relationship between the solid and liquid phases of a material. The solidus line denotes the temperature below which the material is completely solid, while the liquidus line marks the temperature above which the material is entirely liquid. The region between these two lines represents the two-phase (solid-liquid) mixture area, where both solid and liquid phases coexist. (b) A mixed state with a small amount of liquid present. (c) A fully solid state. (d) A fully liquid state. (e) A mixed state with a small amount of solid present.

We utilize the Al-Mg alloy as a prototypical case to elucidate our approach in handling a generic system involving elemental partitioning. The interatomic potential employed is the Al-Mg Finnis–Sinclair type EAM developed by Mendelev et al. [19], which is accessible for download from the NIST potential repository [20,21].

For a specified composition, say Al-10%Mg (atomic percent), the overarching simulation procedure resembles that of a simple system devoid of elemental partitioning, albeit with two notable distinctions. Firstly, contrary to the customary solid-liquid coexistence method that initiates with a balanced mixture of solid and liquid phases (Fig. 1c), our initial configuration comprises a minor fraction of liquid embedded within a solid matrix, as illustrated in Fig. 1d. Secondly, the identification of the nominal melting point deviates from the traditional approach of seeking a mixed state with an equal proportion of solid and liquid. Instead, we target a configuration containing 90% solid and 10% liquid to define this critical temperature. This modification is evident in the fitting results depicted in Fig. 4, where the melting point determination differs significantly from that in Fig. 2. Specifically, the nominal melting point aligns closely with the transition point separating the mixed state from the fully solid state, as seen in Fig. 4b to Fig. 4d. Furthermore, the mixed state in the Al-Mg alloy spans a discernible temperature range, highlighted by the shaded area in Fig. 4d. This contrasts sharply with pure aluminum, where the transition from fully solid to fully liquid occurs abruptly, accompanied by a negligible mixed region, as exemplified in Fig. 2d. Fig. 4e compiles the $T_{min}$, $T_{max}$, and intermediate melting point estimates ($T_m^i$) encountered during the simulations, plotting their evolution against the iteration count (with 0 representing the fast-scanning stage). This visualization underscores both the convergence of the iterative fitting process and the robustness of the final nominal melting point estimate, highlighting the stability and reliability of our approach.

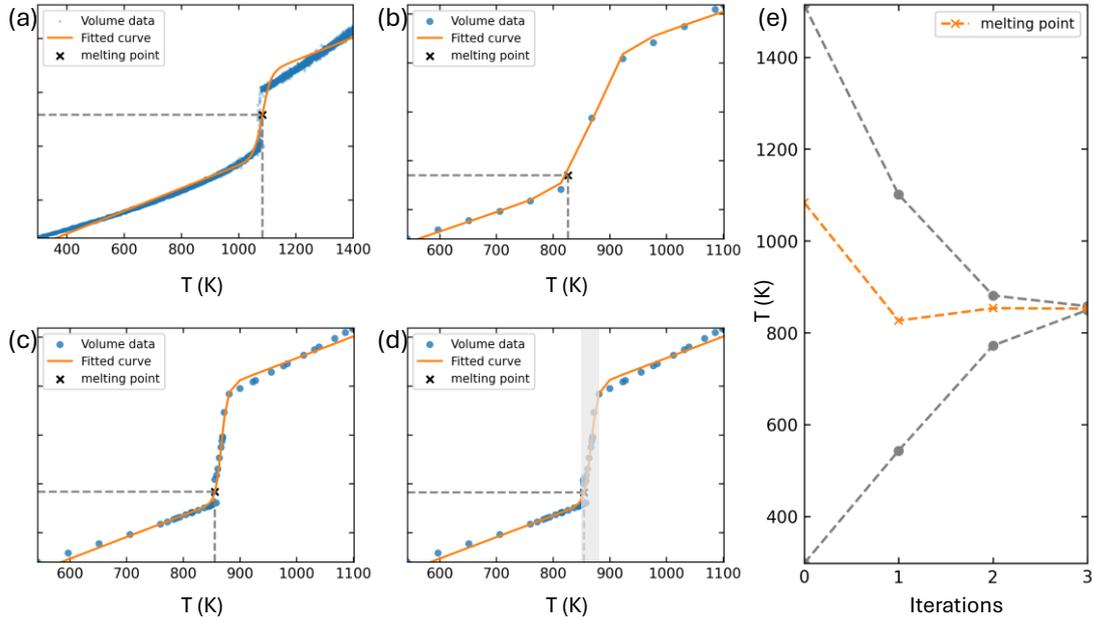

Figure 4. Fitting the *T~V* (temperature versus volume) data of Al-10%Mg (*at*%) to estimate the melting point through iterative processes. (a) The initial *T~V* data obtained from the fast-scanning process. The refined *T~V* data after (b) the first iteration, (c) the second iteration and (d) the third iteration. The shaded region in (d) indicates the solid-liquid coexistent states, showing the continuous changing from solid to liquid. (e) Variation of the estimated melting point, the highest simulating temperature, and the lowest simulating temperature with respect to the number of iterations, illustrating the convergence of the iterative fitting process and the stability of the final melting point estimate.

To obtain a more precise estimation of the nominal melting point and solidification point, the *T~V* data presented in Fig. 4d can undergo further analysis. As depicted in Fig. 5a, the data is segregated into three distinct groups: solid data, mixed data, and liquid data. Subsequently, a linear regression analysis is performed on each of these datasets. The intersection of the regression lines for the solid data and the mixed data represents the melting point. This intersection signifies the equilibrium state where a negligible fraction of liquid exists within a primarily solid matrix, with the solid phase composition reflecting the overall average composition. Analogously, the intersection point between the regression lines of the liquid data and the mixed data defines the solidification point. At this critical juncture, the equilibrium state delicately balances a predominance of liquid with an infinitesimal amount of solid, where the liquid phase composition aligns with the average composition. This refined approach ensures a more accurate determination of both the melting and solidification points.

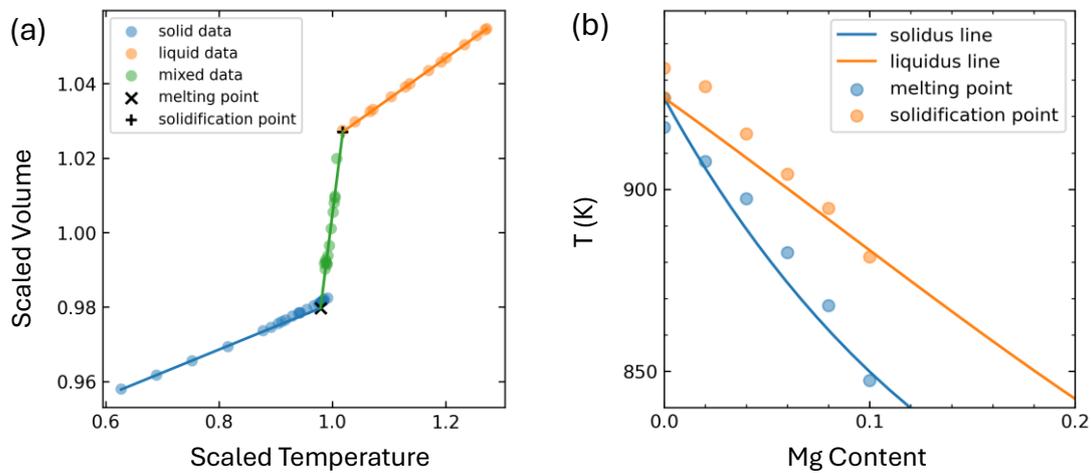

Figure 5: (a) Schematic representation of determining melting and solidification points from $T \sim V$ data of Al-10%Mg (*at*%). The melting point corresponds to the temperature where the solid phase commences melting, transitioning into a mixed phase of solid and liquid. Conversely, the solidification point signifies the temperature at which the liquid phase initiates solidification, forming a mixed phase before fully transitioning to the solid state. The melting point is accurately determined by the intersection of lines that best fit the solid and mixed state data points. Similarly, the solidification point is pinpointed by the intersection of lines that fit the liquid and mixed state data. (b) Comparison of the simulated melting points and thermodynamically derived solidus line, and the simulated solidification points and thermodynamically derived liquidus line.

To validate the reliability of the simulated melting and solidification points, we compared the simulated results with the solidus and liquidus lines derived from a thermodynamics model, as depicted in Fig. 5b. To compute the solidus and liquidus lines, we rely on the Gibbs free energies of both the solid and liquid phases, which are formulated as functions of Mg content ($x$) and temperature ($T$). At a given temperature, the relationship between chemical potential difference $\Delta\mu = \mu_{Mg} - \mu_{Al}$ and Mg content ($x$) can be simulated using the semi-grand canonical Monte Carlo method. In this simulation, the total number of atoms and the chemical potential difference $\Delta\mu$ are kept constant, while the Al and Mg atoms are allowed to switch to achieve equilibrium composition. This relationship can then be represented by the following equation:

$$\Delta\mu(x,T) = \alpha + \beta x + \gamma x^2 + k_B T \ln\left(\frac{x}{1-x}\right) \quad (2)$$

Subsequently, the Gibbs free energy can be integrated, incorporating the Gibbs free energy of pure Al $G_{Al}(T)$, as follows:

$$G(x,T) = G_{Al}(T) + \alpha x + \frac{\beta}{2}x^2 + \frac{\gamma}{3}x^3 + k_B T[x \ln(x) + (1-x)\ln(1-x)] \quad (3)$$

Here, $G_{Al}(T)$ can be integrated using the Gibbs-Helmholtz equation:

$$\frac{\partial(G/T)}{\partial T} = -\frac{H}{T^2} \quad (4)$$

The enthalpy ($H$) of solid and liquid states of Al can be simulated through MD at different temperatures and fitted using quadratic equations:

$$H(T) = H_0 + AT + BT^2 \quad (5)$$

By combining equations (3) and (4), we can express the Gibbs free energy of Al as a function of temperature:

$$G_{Al}(T) = G_0 \frac{T}{T_0} + H_0\left(1 - \frac{T}{T_0}\right) - BT(T - T_0) - AT\ln\left(\frac{T}{T_0}\right) \quad (6)$$

Here, $T_0$ represents the melting point of Al, which is 925 K. With these formulations, we calculate the solidus and liquidus lines, as shown in Fig. 5b. The comparison results demonstrate that our simulated outcomes align well with the thermodynamics model, confirming the validity of our approach in predicting the melting and solidification points of a generic system involving elemental partitioning.

For more complex systems, obtaining the relation between $\Delta\mu$ and composition is nontrivial, *e.g.* for compounds or systems with vacancy. Nevertheless, our method has no limitations and can be readily applied to simulate the melting and sonification points. To demonstrate the efficacy of our approach, we apply it to calculate the solidus and liquidus lines of $HfC_{1-x}$, a highly refractory material renowned for its extreme thermal stability. We employ a Deep Potential model [2] of the NVNMD (Non-Von Neumann Molecular Dynamics) type [22], tailored to harness the computational prowess of Non-Von Neumann architecture hardware, thereby accelerating the simulations by an order of magnitude. This model is rigorously trained using data sourced from a reputable reference [7]. Fig. 6 presents the predicted solidus and liquidus lines, showcasing a notable trend: as the vacancy concentration (*x*) in the C site increases, the melting point initially rises, peaking at a composition range of $x \approx 0.10$ to $0.18$, before subsequently declining. This trend aligns seamlessly with experimental observations [23,24], validating the accuracy of our simulations. Notably, the simulated solidus lines exhibit a steep slope, mirroring the experimental findings, while the liquidus lines follow a more gradual progression. Consistent with previous studies [7], our results indicate that the maximum melting point, hovering close to 4100 K, occurs within the aforementioned vacancy concentration range. This achievement underscores the capability of our method in accurately predicting material properties under varying compositions, thereby offering valuable insights into the behavior of highly refractory materials like $HfC_{1-x}$.

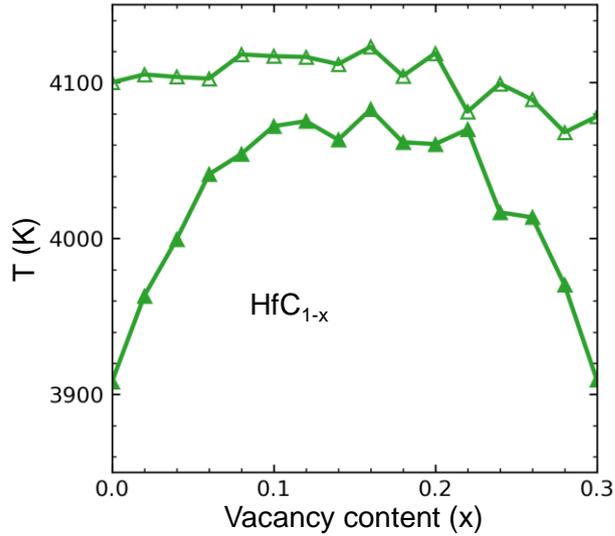

Figure 6. Simulated solidus and liquidus lines of HfC$_{1-x}$ by our method, which agree well with experimental results.

**3.3 Discussion**

When addressing the melting and solidification points, it is crucial to initiate the process with a mixed state containing a small quantity of liquid. This choice stems from the fundamental differences in the diffusivity characteristics between the solid and liquid phases. The solid phase exhibits extremely low diffusivity, often resulting in negligible diffusion events during MD simulations. Consequently, the local composition of the solid remains largely fixed throughout the simulation. In contrast, the liquid phase boasts high diffusivity, facilitating a rapid and uniform distribution of its constituents. This means that any initial non-uniformities in the liquid's composition tend to dissipate quickly, resulting in a more homogeneous state.

To clarify, let's consider the solidification and melting processes of HfC$_{0.7}$ as illustrative examples. Fig. 7 depicts the solidification process of HfC$_{0.7}$, commencing from various initial states, and the subsequent evolution of the composition profile post-solidification. Initially, the composition profile is uniform, as evidenced by the blue lines in Fig. 7a and Fig. 7c, which are generated by randomly removing C atoms from HfC. In the case where the initial state comprises a mixed phase with equal proportions of solid and liquid, upon solidification, the liquid phase undergoes significant depletion of C, while the newly formed solid phase becomes enriched with C. Notably, the composition of the preexisting solid phase remains almost unchanged, as illustrated in Fig. 7a. For a more extreme scenario, where the initial mixed state contains a minimal solid phase, elemental partitioning during solidification leads to a more heterogeneous composition profile, as depicted in Fig. 7c. This indicates that sequential solidification of different regions and the low diffusivity of the solid phase result in a solid phase with non-uniform elemental distribution. Conversely, during the melting process, such issues are mitigated due to the high diffusivity of the liquid phase. Fig. 8 showcases the evolution of composition profiles during the melting of $HfC_{0.7}$. Irrespective of the initial state, the final state converges to a similar configuration, as evident from Fig. 8a and Fig. 8c. Specifically, the composition profile within the liquid phase becomes uniform, closely approximating the average composition when only a small fraction of solid remains.

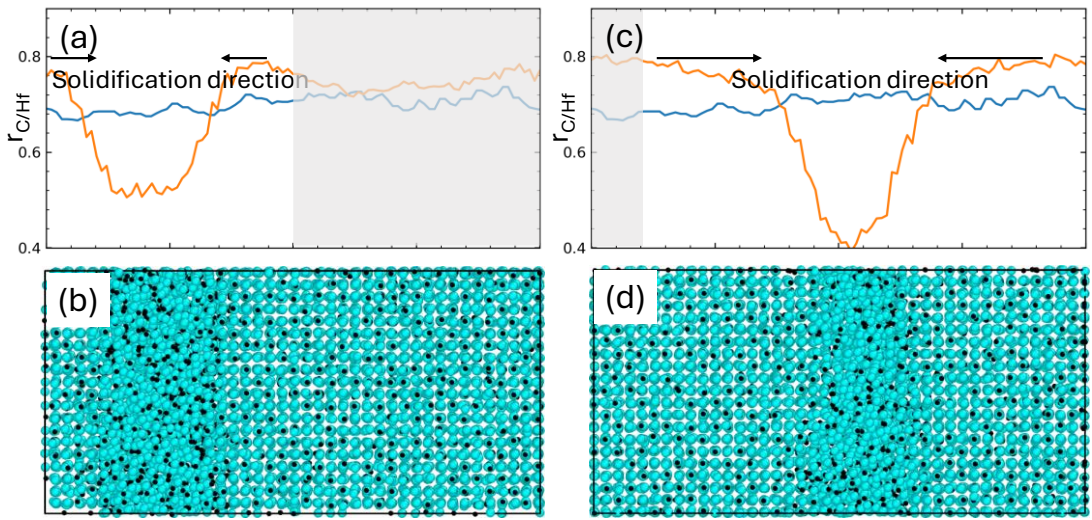

Figure 7. Illustration of elemental partitioning during solidification of $HfC_{0.7}$. (a) and (c) Spatial variation of the C-to-Hf ratio ($r_{C/Hf}$) before and after solidification simulation. The shaded region represents the initial solid phase region of the mixed state. The blue line depicts the initial spatial distribution of $r_{C/Hf}$, indicating a uniform elemental distribution across the system. In contrast, the orange line shows the final spatial variation of $r_{C/Hf}$, highlighting a notable enrichment of C in the solid phase and a corresponding depletion of C in the remaining liquid phase, demonstrating

elemental partitioning during solidification. (b) and (d) The atomic structure after simulation.

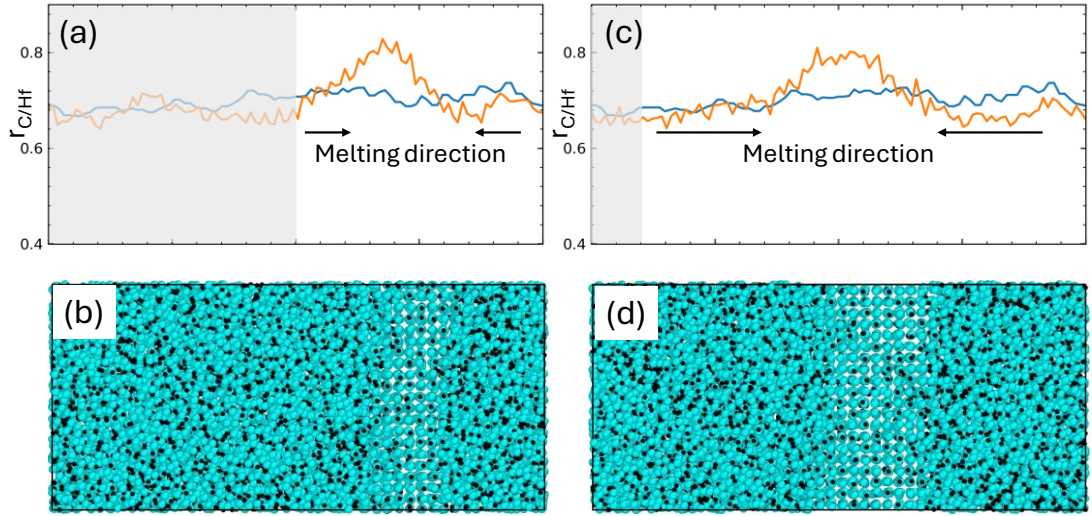

Figure 8. Illustration of elemental partitioning during melting of $HfC_{0.7}$. (a) and (c) Spatial variation of the C-to-Hf ratio ($r_{C/Hf}$) before and after melting simulation. The shaded region represents the initial liquid phase region of the mixed state. The blue line depicts the initial spatial distribution of $r_{C/Hf}$, indicating a uniform elemental distribution across the system. In contrast, the orange line shows the final spatial variation of $r_{C/Hf}$, highlighting a slight enrichment of C in the remaining solid phase and a negligible depletion of C in the liquid phase, demonstrating elemental partitioning during melting. (b) and (d) The atomic structure after simulation.

At the nominal melting point, the equilibrium state primarily consists of a solid matrix with a negligible presence of liquid. Within this matrix, the solid phase composition is uniformly distributed, reflecting the average composition of the material. To prevent the emergence of a non-uniform composition profile, it is advisable to initiate the process with a mixed configuration that contains a minor liquid phase, considering the low diffusivity of the solid phase. By doing so, the majority of the material maintains a uniform composition profile as intended, ensuring that the simulated configuration closely approximates the equilibrium state. Conversely, at the nominal solidification point, the equilibrium state is delicately balanced with a predominant liquid phase and an infinitesimal amount of solid. Here, the liquid phase composition precisely aligns with the average composition of the material. Regardless of the initial mixed state, the composition profile rapidly becomes uniform due to the high diffusivity of the liquid phase, facilitating the achievement of the equilibrium state with ease. Therefore, the initial structure for simulating the nominal melting and solidification point is the one shown in Fig. 1d, which is a mixed state with minor liquid phase.

Contrary to previous methodologies that solely rely on the equilibrium of Gibbs free energy between solid and liquid phases to estimate the melting point, we employ two distinct approaches contingent upon the occurrence of elemental partitioning during the melting process. In the absence of elemental partitioning, we adhere to the conventional approach by searching for the equilibration between the solid and liquid phase. However, in cases where elemental partitioning takes place, we have devised a novel approach to pinpoint critical events. For instance, the nominal melting point signifies the onset of melting from the solid phase, whereas the nominal solidification point marks the initiation of solidification from the liquid phase. Distinguishing between these two points is crucial in practical applications, such as determining the solidification point of an alloy during casting or assessing the risk temperature based on the nominal melting point. In some instances, the disparity between these temperatures can exceed hundreds of Kelvin, as illustrated in Fig. 6. The melting point derived from traditional methods often deviates significantly from either of these temperatures, rendering them less meaningful in practice. Conversely, our methodology offers more insightful information regarding materials' melting and solidification behaviors. Furthermore, given that most materials exhibit elemental partitioning during melting or solidification, our approach holds significant practical value.

In our methodology, we pinpoint critical events by fitting the temperature-volume ($T$~$V$) data, a readily accessible metric in any MD simulation. The prerequisite for this approach is a pronounced deviation in volumes between the solid and liquid phases, a characteristic exhibited by a wide array of materials.

Consequently, our methodology is inherently versatile and can be applied to the majority of materials without restriction. One of the key advantages of our approach lies in its ability to circumvent the need for explicitly identifying the state of each simulation snapshot, such as distinguishing between solid and liquid phases, as previously attempted by Zhu et al. [11]. Accurately and automatically classifying the local structures for any instantaneous configuration during a simulation is a non-trivial task. As a result, their method is largely confined to elemental crystals with simplistic crystal structures. Furthermore, while our primary focus is on fitting $T\sim V$ data, it is noteworthy that other physical quantities, such as potential energy, could potentially offer valuable insights for identifying critical events and could therefore serve as alternative fitting targets, further enhancing the flexibility and applicability of our approach.

## 4. Conclusions

In this study, we present a comprehensive, fully automated workflow for simulating the melting point of materials utilizing MD simulations. Our approach incorporates two distinct simulation modes, tailored to account for the occurrence or absence of elemental partitioning between the solid and liquid phases during melting and solidification.

When the compositions of the solid and liquid phases remain identical, indicating no partitioning, the melting point is defined as the temperature at which both phases coexist in equal proportions. Conversely, in cases where elemental partitioning occurs, we estimate both the nominal melting point and the nominal solidification point. The nominal melting point signifies the temperature threshold at which the solid phase initiates melting, while the nominal solidification point marks the onset of liquid phase solidification.

To determine these critical temperatures, our method relies on fitting the temperature-volume ($T\sim V$) data, a technique applicable to a wide range of materials exhibiting significant volume differences between their solid and liquid states. The simplicity and accessibility of our approach are underscored by its minimal input requirements: solely an interatomic potential file and a sufficiently large supercell representation of the crystal structure. This streamlined workflow facilitates both comprehension and implementation, making it a valuable tool for investigating the melting behavior of diverse materials.

**Code availability**

The code can be accessed from the website

https://bohrium.dp.tech/notebooks/69876147598.

**Declaration of competing interest**

The authors declare that they have no known competing financial interests or personal relationships that could have appeared to influence the work reported in this paper.

**Declaration of generative AI and AI-assisted technologies in the writing process**

During the preparation of this work the author(s) used ERNIE Bot in order to improve language and readability. After using this tool/service, the author(s) reviewed and edited the content as needed and take(s) full responsibility for the content of the publication.

**Acknowledgements**

This work was supported by the National Key Research and Development Program of China (No. 2022YFB3709000) and by the funding provided by University of Science and Technology Beijing. The computing resource of this work was provided by the Bohrium Cloud Platform (https://bohrium.dp.tech), which is supported by DP Technology.